\documentclass[10pt, twocolumn, prl]{revtex4-1}

\usepackage{natbib}

\pdfoutput=1
\usepackage{amsmath,amssymb,epsfig,color,bbold}
\usepackage{graphicx,caption}

\usepackage{bm}
\usepackage{latexsym}
\usepackage{amssymb}
\usepackage{epsfig,amsmath,graphicx}
\usepackage{listings,setspace}

\normalsize

\newcommand{\Lag}{\mathcal{L}}

\usepackage[
	colorlinks=true,
	urlcolor=blue,
	linkcolor=red,
	citecolor=blue
]{hyperref}

\usepackage{tikz}
\usetikzlibrary{arrows,decorations.markings,decorations.pathmorphing}

\begin{document}

\begin{titlepage}

\begin{flushright}
FERMILAB-PUB-17-613-T\\
MITP/17-106

\end{flushright}

\begin{center}
\Large\bf 
Higgs Pair Production as a Signal of Enhanced Yukawa Couplings 
\end{center}

\begin{center}
{\sc   Martin Bauer$^1$, Marcela Carena$^{2,3,4}$, Adri\'an Carmona$^{5}$}\\
\vspace{0.4cm}
{$^1$ Institut f\"ur Theoretische Physik, Universit\"at Heidelberg, Philosophenweg 16, 69120 Heidelberg, Germany,\,\\
$^2$ Fermilab, P.O. Box 500, Batavia, IL 60510, USA,\, \\$^3$ Enrico Fermi Institute, University of Chicago, Chicago, IL 60637, USA, \\$^4$ Kavli Institute for Cosmological Physics,University of Chicago, Chicago, IL 60637, USA\,,  \\$^5$ PRISMA Cluster of Excellence \& Mainz Institute for Theoretical Physics, Johannes Gutenberg University, \\55099 Mainz, Germany
}
\end{center}
\vspace{-0.4cm}
\begin{abstract}
We present a non-trivial correlation between the enhancement of the Higgs-fermion couplings and the Higgs pair production cross section in two Higgs doublet models with a flavour symmetry. This symmetry suppresses flavour-changing neutral couplings of the Higgs boson and allows for a partial explanation of the hierarchy in the Yukawa sector. After taking into account the constraints from electroweak precision measurements, Higgs coupling strength measurements, and unitarity and perturbativity bounds, we identify an interesting region of parameter space leading to enhanced Yukawa couplings as well as enhanced di-Higgs gluon fusion production at the LHC reach.
This effect is visible in both the resonant and non-resonant contributions to the Higgs pair production cross section. We encourage dedicated searches based on differential distributions as a novel way to indirectly probe enhanced Higgs couplings to light fermions.
  \end{abstract}
\maketitle
 
\end{titlepage}
{\bfseries{Introduction.}}\,
Probing the Higgs couplings to the first and second generation fermions is one of the main objectives of the  Higgs program at the LHC. 
New physics could induce very large deviations from the Standard Model (SM) predictions, by changing the way the Higgs couples to light fermions through higher dimensional operators  
\begin{align}\label{eq:l1}
-\mathcal{L}=y_f\bar f \phi f+ y_f' \frac{\phi^\dagger \phi}{\Lambda^2} \bar f \phi f\, +\mathcal{O}\Big(\frac{1}{\Lambda^4}\Big),
\end{align}
where $\phi$ denotes the Higgs doublet and $f$ is an arbitrary fermion. From \eqref{eq:l1} follows for the fermion mass matrix
\begin{align}\label{eq:Mf}
m_f=\left(y_f+y_f'\frac{v^2}{2\Lambda^2}\right)\frac{v}{\sqrt{2}}\,,
\end{align}
while the couplings to the SM Higgs are given by
\begin{align}\label{eq:yf}
g_{hff}=\left(y_f+3y_f'\frac{v^2}{2\Lambda^2}\right)\frac{1}{\sqrt{2}}= \frac{m_f}{v}+\frac{y_{f}'\,v^2}{\sqrt{2}\Lambda^2}\,.
\end{align}
Enhancements of Higgs couplings to light fermions can be induced if the last term in \eqref{eq:yf} becomes sizable with respect to $m_f/v$. This requires fine-tuning between $y_f$ and $y_f'$ in order to recover the observed fermion masses. In addition, \eqref{eq:yf} in general induces sizable flavour-changing neutral currents (FCNCs) mediated by the Higgs, due to the misalignment between fermion masses and Higgs couplings, which requires additional fine-tuning to fulfill the bounds from flavour observables \cite{Goertz:2014qia}. An alignment of the couplings $y_f$ and $y_{f'}$ at a high scale is not stable under renormalization group evolution, because the SM Yukawa coupling and the dimension six operator in \eqref{eq:l1} run differently \cite{Jenkins:2013zja, Jenkins:2013wua, Alonso:2013hga}. \\
In two Higgs doublet models (2HDMs) with a flavour symmetry, the light fermion masses can be explained through higher order operators, avoiding the need for the very small Yukawa couplings in the SM. These higher order operators introduce enhanced diagonal couplings between the Higgs and the SM fermions. Moreover, the structure of Higgs couplings to fermions are close to minimal flavour violating, leading to suppressed FCNCs~\cite{Bauer:2015fxa, Bauer:2015kzy}. 
In this letter we argue that in these models there exists a strong correlation between maximally enhanced Higgs couplings to fermions and a enhanced Higgs pair production that can be probed at the LHC.

Several strategies to test light fermion Yukawa couplings have been proposed, which are sensitive to enhanced couplings present in the class of models discussed in this letter. In the case of muon and electron Yukawa couplings,
direct measurements of $h\rightarrow \mu^+\mu^-$ and $h\rightarrow e^+e^-$ yield the strongest constraints \cite{Khachatryan:2014aep, Aad:2014xva, Khachatryan:2016vau}
\begin{align}
|\kappa_\mu| < 2.1\,,\quad |\kappa_e|\lesssim 608\,,
\end{align}
where $\kappa_f=g_{hff}/g^\text{SM}_{hff}$. 
Direct measurements of the Higgs couplings to light colored fermions are much more challenging. The strongest (yet indirect) bounds follow from a combined fit to Higgs coupling strength measurements, allowing only one Yukawa to deviate at a time~\cite{Yu:2016rvv}
\begin{align}
|\kappa_d|<1270\,,\quad |\kappa_u|<1150\,,\quad |\kappa_s|<53\,\quad |\kappa_c|<5\,.
\end{align}
More model-independent methods are inclusive measurements of $h\rightarrow c\bar c$ or associated production of $pp \rightarrow hc + h\bar c$, which  strongly depend on the $c-$ and $b-$tagging efficiencies \cite{Perez:2015lra, Perez:2015aoa,Brivio:2015fxa}. Exclusive, radiative Higgs decays $h\rightarrow J/\psi (\Upsilon) \gamma$ provide an alternative way to test charm (and bottom) Yukawas and notably also to access their sign \cite{Isidori:2013cla, Bodwin:2013gca,Kagan:2014ila,Koenig:2015pha}. Measurements of the total width of the Higgs offer another handle on individual Yukawa couplings \cite{Perez:2015aoa}, as well as measurements of $p_T-$distributions in $pp\rightarrow h$ and $pp\rightarrow hj$  \cite{Bishara:2016jga, Cohen:2017rsk}.
A novel strategy based on measuring the charge asymmetry in $W^\pm h$ has been proposed \cite{Yu:2016rvv}. If Higgs couplings to proton valence quarks and electrons are simultaneously enhanced, even frequencies of atomic clocks could be modified \cite{Delaunay:2016brc}. \\[-.6cm]

{\bfseries{\scshape{Formalism.}}}\,\,
In 2HDMs, the SM singlet operator $\phi_1\phi_2$ can carry a flavour charge, such that for a given flavour the SM Yukawa coupling is replaced by a higher order operator 
\begin{align}\label{eq:rep}
y_f\bar f_L \phi f_R \quad \rightarrow\quad y_f' \left(\frac{\phi_1 \phi_2}{\Lambda^2}\right)^{n_f} \bar f_L \phi_i f_R\,,
\end{align}
in which $\Lambda$ is the suppression scale, $\phi_i$ is either $\phi_1$ or $\phi_2$ and the integer $n_f$ depends on the flavor charge assigned to $f_L \phi_i f_R$ and $\phi_1\phi_2$.
As a consequence, the corresponding fermion masses are given by
\begin{align}\label{eq:eps}
m_f=y_{f}'\,\varepsilon^{n_f} \frac{v}{\sqrt{2}}\,, \quad \varepsilon = \frac{v_1 v_2}{2\Lambda^2}=\frac{t_\beta}{1+t_\beta^2}\frac{v^2}{2\Lambda^2},
\end{align}
with the vacuum expectation values $\langle\phi_{1, 2}\rangle=v_{1, 2}$ and $t_\beta \equiv v_1/v_2$.  
For the right choice of flavor charges, the hierarchy of SM fermion masses and mixing angles can be explained by higher order operators \cite{Bauer:2015fxa, Bauer:2015kzy}. In contrast to the ansatz \eqref{eq:l1}, lower dimensional operators can be forbidden by these flavor charges.
In the following, we will illustrate our result based on the Lagrangian
\begin{align}\label{eq:yuksI}
\Lag_Y^\mathrm{I}&\ni  y^u_{ij}\, \bigg(\!\frac{\phi_1^{\phantom{\dagger}} \phi_2^{\phantom{\dagger}}  }{\Lambda^2}\!\bigg)^{\!n_{u_{ij}}}\! \bar Q_i \phi_1\, u_j + y^d_{ij}\,\bigg(\!\frac{\phi_1^\dagger \phi_2^\dagger }{\Lambda^2}\!\bigg)^{\!n_{d_{ij}}}\!\bar Q_i \tilde \phi_1\, d_j\notag\\
&+  y^\ell_{ij}\,\bigg(\!\frac{\phi_1^\dagger \phi_2^\dagger }{\Lambda^2}\!\bigg)^{\!n_{\ell_{ij}}}\!\bar L_i \tilde \phi_1\, \ell_j +h.c.\,,
\end{align}
which reduces to a 2HDM of type I in the limit $n_u, n_d, n_\ell \rightarrow 0$. This expression can be readily extended to other types of 2HDMs \cite{Bauer:2015kzy} and the discussion in the remainder of the paper holds independent of this choice.
The Higgs sector contains two neutral scalar mass eigenstates $h, H$, one pseudoscalar $A$ and one charged scalar $H^\pm$ and we identify the lighter scalar mass eigenstate $h$ with the $125$ GeV resonance observed at the LHC. 
The couplings between the scalars and the electroweak gauge bosons are fixed as in any 2HDM to $g_{\varphi VV}=\kappa^\varphi_V \, 2\,m_V^2/v$, with
$\kappa^h_V=s_{\beta-\alpha}\,, \, \kappa^H_V=c_{\beta-\alpha}$ for $V=W^\pm, Z$, and we use the notation $s_x=\sin (x)$, $c_x=\cos(x)$ and $t_x=\tan(x)$.
The couplings between the scalars $\varphi=h,H$ and SM fermions $f_{L_i, R_i}= P_{L,R} f_i$ in the mass eigenbasis read \\[-.4cm]
\begin{equation}\label{eq:newlag}
\Lag= g_{\varphi f_{L_i} f_{R_j} }\, \varphi \,\bar f_{L_i} f_{R_j}+h.c.
\end{equation}
with 
a flavor index $i$, such that $u_i=u, c,t$, $d_i=d,s,b$ and $\ell_i=e,\mu,\tau$. This induces 
flavor-diagonal couplings 
\begin{align}\label{eq:diagcoup}
\!\!\! \! g_{\varphi f_{L_i}f_{R_i}}\!= \!\kappa^\varphi_{f_i}\, \frac{m_{f_i}}{v}\!=\!\Big(g^{\varphi}_{f_i}(\alpha,\beta)+n_{f_i}\, f^\varphi(\alpha, \beta)\!\Big)\frac{m_{f_i}}{v},
\end{align}
and flavor off-diagonal couplings
 \begin{align}\label{eq:foff}
g_{\varphi f_{L_i}f_{R_j}}&=  f^\varphi(\alpha, \beta)\Big(\mathcal{A}_{ij}\frac{m_{f_j}}{v}-\frac{m_{f_i}}{v}\mathcal{B}_{ij} \Big)\,.
\end{align}
The flavor universal functions in \eqref{eq:diagcoup} and \eqref{eq:foff} are given by
\begin{align}
g^h_{f_i}=\frac{c_{\beta-\alpha}}{t_\beta}+s_{\beta-\alpha}\,,\qquad g^H_{f_i}=c_{\beta-\alpha}-\frac{s_{\beta-\alpha}}{t_\beta}\,,
\end{align}
and
\begin{align}\label{eq:fFs}
f^h(\alpha,\beta)&=c_{\beta-\alpha}\Big(\frac{1}{t_\beta}-t_\beta\Big)+2s_{\beta-\alpha}\,,\\ 
f^H(\alpha,\beta)&=-s_{\beta-\alpha}\Big(\frac{1}{t_\beta}-t_\beta\Big)+2c_{\beta-\alpha}\,.
\end{align}
Flavor off-diagonal couplings between the neutral scalars and SM fermions are induced in $\eqref{eq:foff}$ through the matrices in flavor space $\mathcal{A}$ and $\mathcal{B}$, whose entries are proportional to the flavor charges of the corresponding fermions that define the coefficients in \eqref{eq:yuksI}.  In general, there are flavor charges of the fermion singlets, $a_{f_i}$,  doublets, $a_{Q_i}$ and $a_{L_i}$, as well as those of the Higgs doublets $a_1$ and $a_2$. We set the flavor charge of $\phi_1\phi_2$ to $a_1+a_2=1$ by fixing $a_2=1$ and $a_1=0$, such that
\begin{align}
n_{u_{ij}}\!\!=a_{Q_i}\!-a_{u_j},\,\,
n_{d_{ij}}\!\!=a_{Q_i}\!-a_{d_j},\,\,
n_{\ell_{ij}}\!\!=a_{L_i}\!-a_{\ell_j}.
\end{align}
While these exponents depend on the relative charge assignments for the two Higgs doublets, the structure of 
the matrices $\mathcal{A}$ and $\mathcal{B}$ is independent of this choice. If all flavor charges for a given type of fermions are equal, the off-diagonal elements of these matrices vanish. Otherwise, for couplings of the neutral scalars to up-type quarks $\mathcal{B}=\mathcal{U}$ with off-diagonal elements
\begin{align}\label{eq:fviol}
\mathcal{U}_{12}&\approx (1\!-\!\delta_{a_{u_1}\!a_{u_2}}\!)\epsilon^{|a_{u_1}\!-a_{u_2}\!|}+ \delta_{a_{u_1}\!a_{u_2}} \epsilon^{|a_{u_3}\!-a_{u_2}\!|+|a_{u_3}\!-a_{u_1}\!|}\,,\notag\\
\mathcal{U}_{13}&\approx (1\!-\!\delta_{a_{u_1}\!a_{u_3}}\!)\epsilon^{|a_{u_1}\!-a_{u_3}\!|}+ \delta_{a_{u_1}\!a_{u_3}} \epsilon^{|a_{u_2}\!-a_{u_1}\!|+|a_{u_2}\!-a_{u_3}\!|}\,,\notag\\
\mathcal{U}_{23}&\approx (1\!-\!\delta_{a_{u_2}\!a_{u_3}}\!)\epsilon^{|a_{u_2}\!-a_{u_3}\!|}+ \delta_{a_{u_2}\!a_{u_3}} \epsilon^{|a_{u_1}\!-a_{u_2}\!|+|a_{u_1}\!-a_{u_3}\!|}\,,
\end{align}\\[-.3cm]
and the same expressions hold for $\mathcal{A}=\mathcal{Q}$ with $a_{u_i}\rightarrow a_{Q_i}$. For couplings between the neutral scalars and down-type quarks $\mathcal{A}=\mathcal{Q}$ and $\mathcal{B}=\mathcal{D}$, where the elements of $\mathcal{D} $ are given by \eqref{eq:fviol} for $a_{u_i}\rightarrow a_{d_i}$. Finally, flavor off-diagonal couplings between charged leptons and neutral scalars are given by \eqref{eq:foff} with 
$\mathcal{A}=\mathcal{C}$ with the elements \eqref{eq:fviol} for $a_{u_i}\rightarrow a_{L_i}$, and $\mathcal{B}=\mathcal{E}$ with the elements \eqref{eq:fviol} for $a_{u_i}\rightarrow a_{\ell_i}$. These structures lead to flavor-FCNCs, which are chirally suppressed and proportional to powers of the ratio~$\varepsilon$. 
The flavor symmetry strongly constrains the Higgs potential 
\begin{align}
V&=\mu_1^2 \phi_1^\dagger \phi_1 + \mu_2^2 \phi_2^\dagger \phi_2 + \left ( \mu_3^2  \phi_1^\dagger \phi_2 + h.c. \right )\notag\\
& + \lambda_1  \hspace{0.25mm} \big ( \phi_1^\dagger \phi_1  \big )^2  + \lambda_2  \hspace{0.25mm} \big ( \phi_2^\dagger \phi_2 \big  )^2 \notag\\
 \phantom{xx}& +  \lambda_3 \hspace{0.25mm} \big ( \phi_1^\dagger \phi_1  \big ) \big ( \phi_2^\dagger \phi_2  \big ) + \lambda_4  \hspace{0.25mm} \big ( \phi_1^\dagger \phi_2  \big ) \big ( \phi_2^\dagger \phi_1  \big )  \,.
\end{align}
The seven independent parameters $\mu_1^2,\mu_2^2,\mu_3^2$ and $\lambda_1, \lambda_2, \lambda_3, \lambda_4$ can be exchanged for the vacuum expectation values $v_1$ and $v_2$,  the physical masses $m_h, M_H, M_A, M_{H^\pm}$ and the mixing angle $c_{\beta-\alpha}$. The coupling between the heavy scalar $H$ and two SM Higgs scalars $h$, as well as the triple Higgs coupling can be expressed as \cite{Boudjema:2001ii, Gunion:2002zf}\vspace{-.3cm}
\begin{align}
&g_{Hhh}=\label{eq:main1}\\
&\quad\frac{c_{\beta-\alpha}}{v}\!\left[\big(1\!-\!f^h(\alpha,\beta)s_{\beta-\alpha}\big)\big(3M_A^2\!-\!2m_h^2\!-\!M_H^2\big)\!-\!M_A^2\right],\notag\\
&g_{hhh}= -\frac{3}{v}\!\left[f^h(\alpha,\beta)c_{\beta-\alpha}^2(m_h^2-M_A^2)+m_h^2s_{\beta-\alpha}\right].\label{eq:main2}
\end{align}
 \begin{figure}
 \includegraphics[width=.5\textwidth]{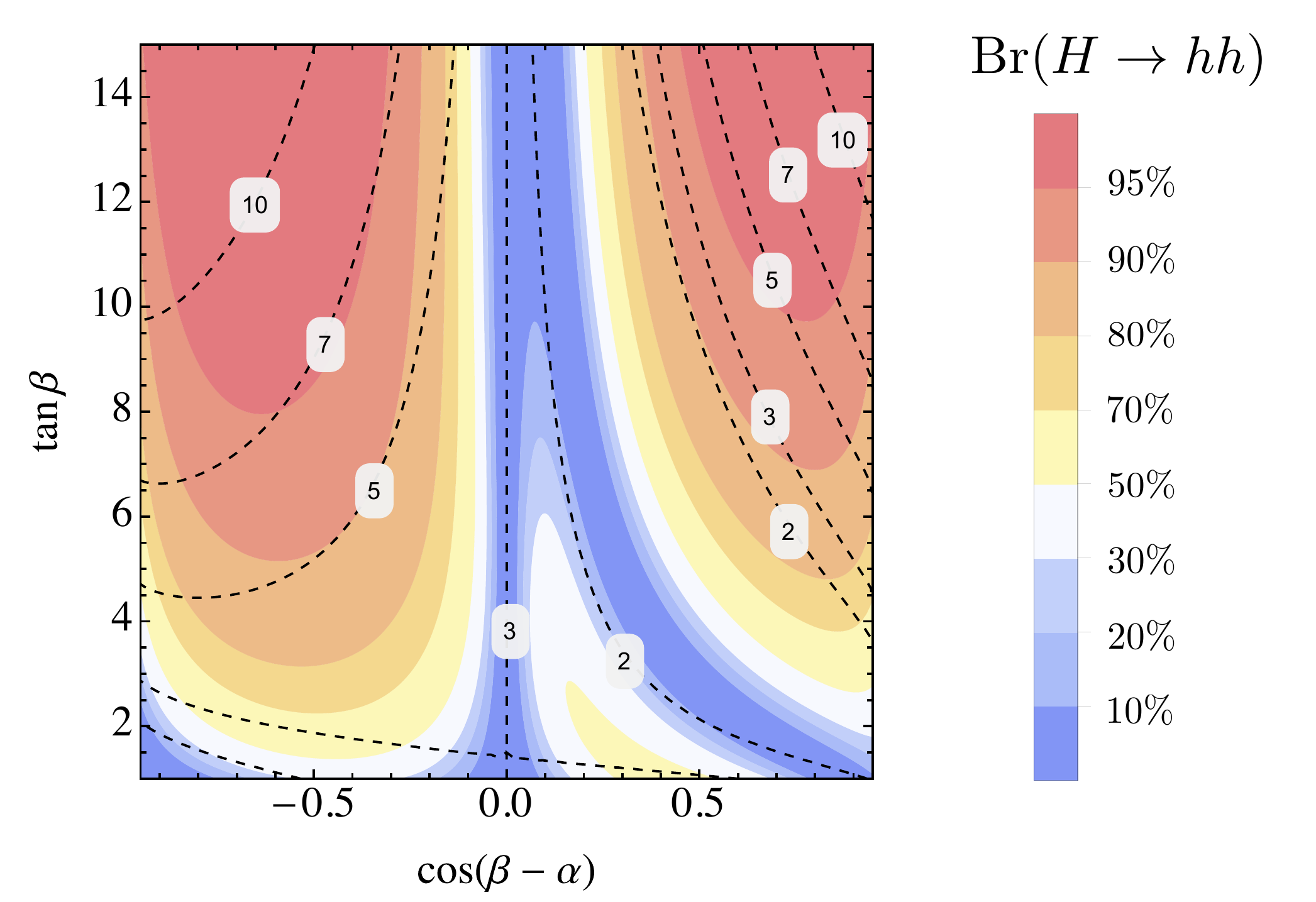}
 \caption{\label{fig:BRvskap} The color coding shows the dependence of $\text{Br}(H \to hh)$ on $c_{\beta-\alpha} $ and $ t_\beta$ for $M_H=M_{H^\pm}
 550$ GeV, $M_A=450$ GeV. The dashed contours correspond to constant $|\kappa_f^h|$ for $n_f=1$.}
  \vspace{-.6cm}
 \end{figure}
\begin{figure*}
\includegraphics[width=.465\textwidth]{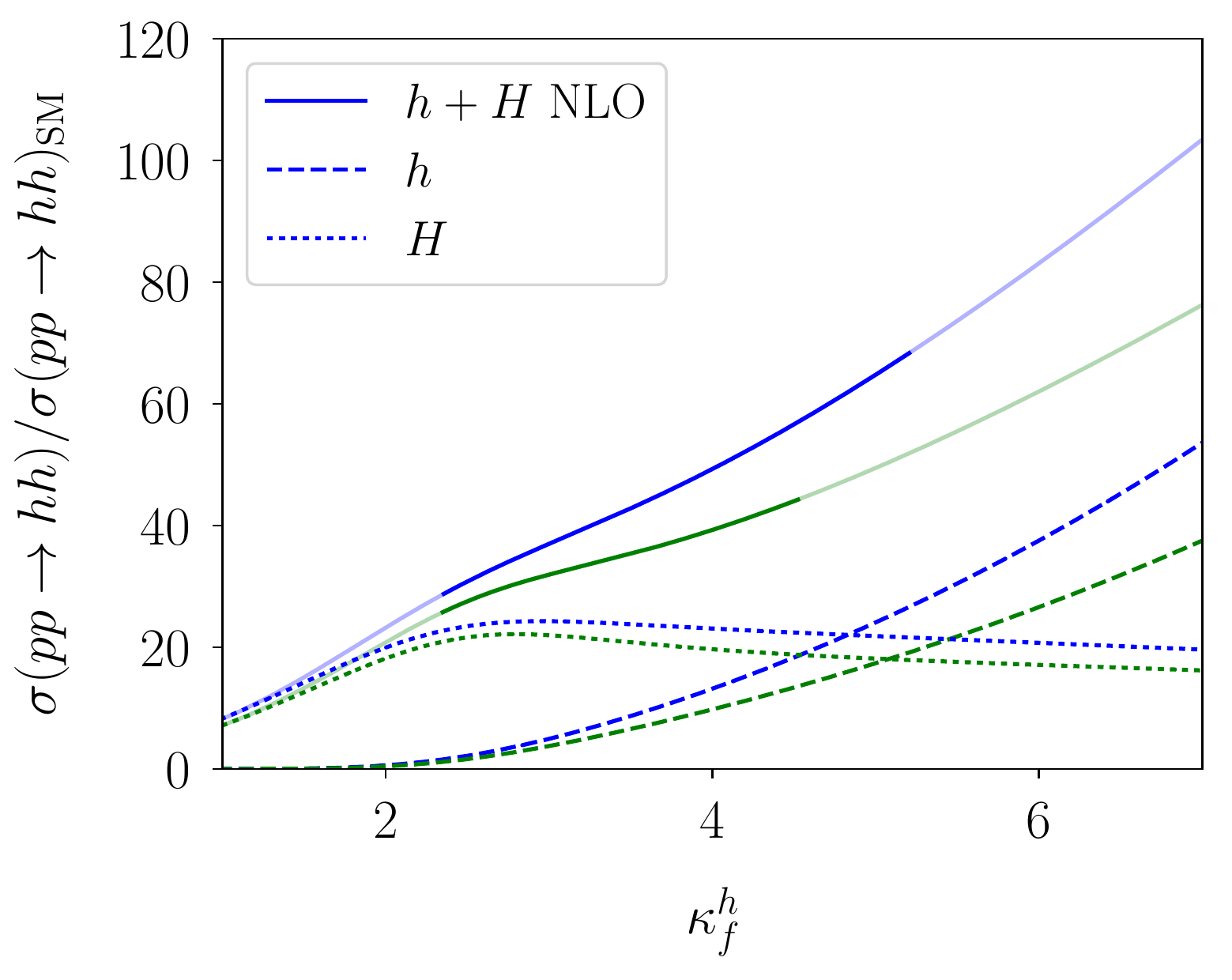}\includegraphics[width=.49\textwidth]{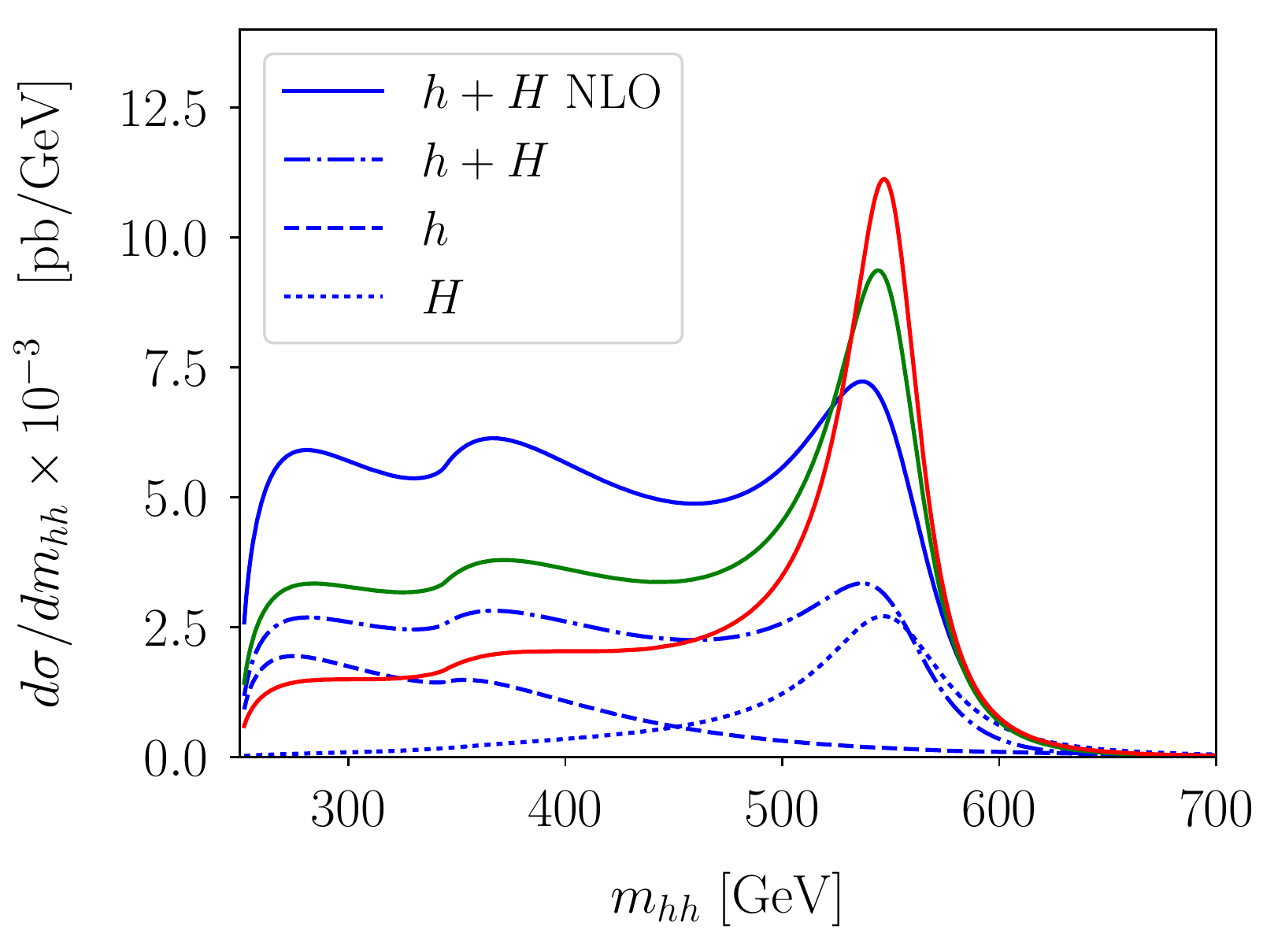}
	\caption{\label{fig:xsec} Left: Cross section for Higgs pair production in units of the SM prediction as a function of $\kappa_f^h$ for $c_{\beta-\alpha}=-0.45~(-0.4)$ and $M_H=M_{H^\pm}=550$ GeV, $M_A=450$ GeV in blue (green) at $\sqrt{s}=13 $ TeV. Right: Invariant mass distribution for the different contributions to the signal with $c_{\beta-\alpha}=-0.45$ and $\kappa^h_f=5$ (blue),  $\kappa_f^h=4$ (green) and $\kappa_f^h=3$ (red) at $\sqrt{s}=13 $ TeV, respectively.} 
\end{figure*}
\textbf{Higgs Pair Production.} The main finding of our paper is that the parameter space for which the diagonal couplings of the SM Higgs to fermions \eqref{eq:diagcoup} are maximally enhanced is directly correlated with an enhancement of the trilinear couplings \eqref{eq:main1} and \eqref{eq:main2}. This parameter space can be identified with the region for which $f^h(\alpha,\beta)\gg 1$, outside of the decoupling limit $c_{\beta-\alpha}=0$. For maximally enhanced couplings, the mass of the heavy scalar $H$ cannot be arbitrarily large and resonant Higgs pair production is a signal of this model. The correlation between the enhancement of the Higgs couplings to SM fermions $\kappa^h_f$ and $\text{Br}(H \to hh)$ is illustrated for $M_H=M_A=M_{H^\pm}=500$ GeV in Fig. \ref{fig:BRvskap}. The color coding shows the dependence of $\text{Br}(H \to hh)$ on $c_{\beta-\alpha} $ and $ t_\beta$, and the dashed contours correspond to constant $|\kappa_f^h|$ for $n_f=1$. This correlation is independent of the factor $n_f$ while $n_f > 1$ leads to larger enhancement factors, and holds throughout the parameter space, apart from the limits $c_{\beta-\alpha}\approx 0$ and $c_{\beta-\alpha}\approx \pm 1$. The latter case is strongly disfavoured by SM Higgs coupling strength measurements, and the correlation breaks down due to the factor $s_{\beta-\alpha}$ in front of $f^h(\alpha,\beta)$ in \eqref{eq:main1}. The limit $c_{\beta-\alpha}=0$  is usually associated with the decoupling of the heavy scalar states, for which $g_{hhh}=-3m_h^2/v$  takes on its SM value and $g_{Hhh}=0$, while the enhancement of Higgs couplings to fermions is fixed to $\kappa_{f_i}^h= 2n_{f_i}+1$. The decoupling limit corresponds to a large value of the pseudoscalar mass $M_A\gg v$, which is related to the spurion 
$\mu_3\propto M_A$ that softly breaks the flavour symmetry assumed in \eqref{eq:yuksI}. At one-loop, one expects this spurion to break the structure of the matrices \eqref{eq:fviol}, inducing FCNCs proportional to $ \mu_3^2/(4\pi \Lambda)^2$. Therefore, the relations we present only hold if additional scalars are present below the TeV scale, for which the parameter space $c_{\beta-\alpha}\neq 0$ is allowed.\\
For larger values of $t_\beta$ there is a suppression of gluon-fusion production, $\sigma(gg\to H)\propto 1+1/t_\beta^2-(\kappa_t^h)^2$, where $\kappa_t^h \approx 1$, that partially cancels the enhancement of $\text{Br}(H \to hh)$. 
However, since $\sigma(gg \to h)\propto (\kappa_t^h)^2$, the cross section $\sigma(gg\to h\to hh)$ is unsuppressed for large values of $t_\beta$ resulting in a continuous correlation between $\kappa_f^h$ and $\sigma(gg\to hh)$ due to the non-trivial interplay between the resonant and non-resonant Higgs pair production processes.  We illustrate this result in the left panel of Fig.~\ref{fig:xsec}, in which the dotted (dashed) lines correspond to the contribution from resonant (non-resonant) Higgs pair production in gluon fusion. 
The solid line is the full $\sigma(gg\to hh)$ in the 2HDM in units of the SM value.
We set $M_{H^\pm}=M_H=550$ GeV, $M_A=450$ GeV, and show values of $c_{\beta-\alpha}= -0.45 (-0.4)$ in green (blue) lines. Higgs coupling strengths measurements and electroweak precision measurements constrain large values of $c_{\beta-\alpha}$, but do not exclude the values considered here for a Yukawa sector of a 2HDM of type I. 
In order to produce the signal, we use our own C\texttt{++} implementation of the NLO QCD cross section for di-Higgs production in the presence of a scalar singlet~\cite{Dawson:2015haa}, in the approximation where the exact $m_t$-dependent form factors are inserted into the $m_t\to \infty$ NLO calculation \cite{Dawson:1998py}. Since the pseudoscalar and the charged Higgs do not contribute, these results can be easily applied here. We use the CT14NLO PDF from LHAPDF6 \cite{Buckley:2014ana} as well as the C\texttt{++} library QCDLoop~\cite{Carrazza:2016gav} to evaluate the corresponding one-loop integrals, neglecting small corrections from quark initial states. Solid lines show the NLO results, while the solid shaded lines mark the values of $\kappa_f$ excluded by perturbativity and unitarity constraints~\cite{Eriksson:2009ws}. The dotted (dashed) lines show the LO ratios for the resonant (non-resonant) contribution. However, to a very good approximation the NLO corrections factorize and drop out of the ratio. \\
For the values of $\kappa_f^h$ considered, $\sigma(pp \to hh)$ never exceeds the experimental bound on the non-resonant Higgs pair production cross section \cite{CMS:2017orf}. The values of $\kappa_f^h$ in Fig.~\ref{fig:xsec} follow from fixing $n_f=1$ and values of $\mathcal{O}(10)$ and larger are obtained for $n_f>1$. 
Note that the correlation between $\sigma (pp \to hh)$ and $\kappa_f^h$ is stronger for vector boson fusion production, because there is no suppression of $\sigma(pp \to H)$ for $t_\beta >1$ and $\sigma(qq\to qqH)\propto s_{\beta-\alpha}^2$.
In the right panel of Fig.~\ref{fig:xsec}, the invariant mass distribution for the different contributions to the signal 
with $c_{\beta-\alpha}=-0.45$ are shown for three values of $\kappa_f^h$ and $\sqrt{s}=13 $ TeV. As a consequence of the enhancement of Higgs-fermion couplings, both non-resonant and resonant contributions are enhanced.  The relevance of the $d\sigma/dm_{hh}$ distribution for both resonant \cite{Baur:2003gp} and non-resonant contributions \cite{Baur:2002rb} to the Higgs pair production cross-section has long been emphasized \cite{Dolan:2012ac,Dolan:2015zja, Kling:2016lay}. Searches for resonant di-Higgs production are sensitive to a peak in the spectrum, which roughly excludes heavy scalar masses $M_H \lesssim 500$ GeV, independent of $f^h(\alpha,\beta)$~\cite{Sirunyan:2017tqo}. For larger $M_H$ and sizable $\kappa_f^h$, the interference between the different contributions turns the broad resonance peak into a shoulder in the $d\sigma/dm_{hh}$ distribution for the total cross section, as shown by the blue line in the right panel of Fig.~\ref{fig:xsec}. Whether current experimental resonance searches can resolve such a structure strongly depends on the shape of the invariant mass distribution~\cite{private}. We encourage a dedicated analysis considering the corresponding $d\sigma/dm_{hh}$ templates to maximize the sensitivity to features in the di-Higgs invariant mass distribution from the simultaneous enhancement of  $g_{hhh}, g_{Hhh}$ and $\kappa_h^f$.\\

\textbf{An Explicit Example. } \,
We now consider a concrete example for which the flavour charges of down-type quarks and leptons vanish $n_{\ell_i}=n_{d_i}=0 \,\,\forall\, i$, whereas the up quarks carry charges $n_t=0, n_c=1, n_u=3$ and we choose all charges of the $SU(2)_L$ fermion doublets to be zero. As a consequence, the top coupling to the SM Higgs $h$ is unchanged from its value in the 2HDM of type I, while charm and up-quark couplings vary with $t_\beta$ and $c_{\beta-\alpha}$ according to \eqref{eq:diagcoup}. This leads to flavour-changing couplings of the SM Higgs to up-type quarks suppressed by powers of the ratio $\varepsilon$,
\begin{align}\label{eq:Umatrix}
\mathcal{U}=\begin{pmatrix}1&\varepsilon^2&\varepsilon^3\\
\varepsilon^2&1&\varepsilon\\
\varepsilon^3&\varepsilon&1\end{pmatrix}\,,\qquad \mathcal{Q}=0\,.
\end{align}
In the up-sector, the strongest constraints on FCNCs arise from $D-\bar D$ mixing. 
Due to the structure of \eqref{eq:foff},  the leading contribution to the Wilson coefficients entering $D-\bar D$ mixing are chirally suppressed and proportional to $\mathcal{U}_{12}^2= \varepsilon^4$. Assuming order one dimensionless coefficients, the experimental limit leads to the constraint \cite{Bona:2007vi}
\begin{align}
\text{Im}\,\bigg(\frac{f^h(\alpha,\beta)}{m_h}\frac{m_c}{v}\,\varepsilon^2\bigg)^2 \lesssim 2\cdot 10^{-14}\,,
\end{align}
where the less relevant contributions from the heavy scalars have been neglected. For  
the maximal values of $f^h(\alpha,\beta)\approx 10$, this yields $\varepsilon \lesssim 1/55 $. This example would lead to a Higgs pair production cross section of $\sigma(pp \to hh)\approx 50\times\sigma_\text{SM}(pp\to hh)$ with enhancements of the Higgs couplings to up-quarks of $\kappa^h_u=10.2$ and to charm-quarks of $\kappa^h_c=4$, respectively. In principle, similar models can be build with flavor charged leptons and down-type quarks. The simultaneous enhancement of $\kappa_\tau$ or $\kappa_b$ and stronger flavor constraints lead to a more constrained parameter space for such models.
  
\textbf{Conclusions. }\,
We report a non-trivial correlation between an enhancement of Higgs couplings to light fermions and enhanced resonant and non-resonant contributions to the Higgs pair production cross section. Such a correlation appears naturally in a class of models in which Higgs-mediated flavour changing currents are suppressed by a flavour symmetry.  We show that even after imposing perturbativity and unitarity bounds as well as constraints from Higgs couplings strength measurements, the parameter space allowing for maximally enhanced Higgs-fermion couplings entails a Higgs pair production cross section exceeding the SM prediction by more than an order of magnitude. Present searches partly probe this interesting correlation, but dedicated LHC studies are required to ultimately explore this idea and indirectly constrain signals of new physics modifying light fermion Yukawa couplings. \\
 
\emph{Acknowledgments}
We thank Luca Cadamuro, Jacobo Konigsberg and Zhen Liu for helpful discussions.
This manuscript has been authored by Fermi Research Alliance, LLC under Contract No. DE-AC02-07CH11359 with the U.S. Department of Energy, Office of Science, Office of High Energy Physics. The United States Government retains and the publisher, by accepting the article for publication, acknowledges that the United States Government retains a non-exclusive, paid-up, irrevocable, world-wide license to publish or reproduce the published form of this manuscript, or allow others to do so, for United States Government purposes. MB acknowledges support from the German Research Foundation (DFG) through the Forschergruppe "New Physics at the Large Hadron Collider" (FOR 2239). The research of Adrian Carmona has been supported by the Cluster of Excellence {\em Precision Physics, Fundamental Interactions and Structure of Matter\/} (PRISMA--EXC 1098) and grant 05H12UME of the German Federal Ministry for Education and Research (BMBF).\\[-1.cm]

\end{document}